\documentclass[aps,prl,twocolumn,showpacs,floatfix]{revtex4}
\usepackage{graphicx}

\begin{document}

\title{Proton Electrodynamics in Liquid Water}

\author{A. A. Volkov$^{1}\footnote {electronic address: aavol@bk.ru}$, V. G. Artemov$^{1}$, A. V. Pronin$^{1, 2}$}

\address{$^{1}$ A. M. Prokhorov Institute of General Physics,
RAS, 119991 Moscow, Russia \\
$^{2}$ Dresden High Magnetic Field Laboratory, HZ
Dresden-Rossendorf, 01314 Dresden, Germany}

\date{December 12, 2012}

\begin{abstract}

The dielectric spectrum of liquid water, $10^{4} - 10^{11}$ Hz, is
interpreted in terms of diffusion of charges, formed as a result of
self-ionization of H$_{2}$O molecules. This approach explains the
Debye relaxation and the dc conductivity as two manifestations of
this diffusion. The Debye relaxation is due to the charge diffusion
with a fast recombination rate, $1/\tau_{2}$, while the dc
conductivity is a manifestation of the diffusion with a much slower
recombination rate, $1/\tau_{1}$. Applying a simple model based on
Brownian-like diffusion, we find $\tau_{2} \simeq 10^{-11}$ s and
$\tau_{1} \simeq 10^{-6}$ s, and the concentrations of the charge
carriers, involved in each of the two processes, $N_{2} \simeq 5
\times 10^{26}$ m$^{-3}$ and $N_{1} \simeq 10^{14}$ m$^{-3}$.
Further, we relate $N_{2}$ and $N_{1}$ to the total concentration of
H$_{3}$O$^{+}$--OH$^{-}$ pairs and to the pH index, respectively,
and find the lifetime of a single water molecule, $\tau_{0} \simeq
10^{-9}$ s. Finally, we show that the high permittivity of water
results mostly from flickering of separated charges, rather than
from reorientations of intact molecular dipoles.

\end{abstract}


\maketitle

\section{Introduction}

Electrical properties of water are of high importance in many vital,
environmental, and technological processes \cite{Robinson, Trimble}.
They have been under intensive investigation for many decades
\cite{Eisenberg, Franks, Hippel, Chaplin}. It has long been
recognized that at room temperature, water is a good insulator with
negligible electronic conductivity and a dielectric constant
$\varepsilon \simeq 80$. Potentiometric measurements reveal an
appreciable proton conductivity, $\sigma_{dc}$ = $5.5 \times
10^{-6}$ $\Omega^{-1}$m$^{-1}$ at room temperature \cite{Light}.
This value is associated with the pH index, which is a key indicator
of activity of protons in chemical reactions \cite{Bates, Bockris}.
Normally, pH = 7; this water is regarded as neutral with the
``free"-proton concentration of $6 \times 10^{19}$ m$^{-3}$. It is
believed that under normal conditions, a given H$_{2}$O molecule
will on average dissociate in roughly 10$^{4}$ seconds (11 hours)
\cite{Geissler, Bakker}.

The origin of the high static dielectric constant of water is
commonly explained by the orientational motion of the molecular
dipoles, which is referenced as the Debye relaxation \cite{Hippel}.
The relaxation is particularly evident in the frequency spectrum of
dielectric permittivity, $\varepsilon^{*}(\omega) = \varepsilon
'(\omega ) + i\varepsilon''(\omega)$, as a strong anomaly near a
characteristic frequency $\nu_{0} = \omega_{0}/2\pi$; $\nu_{0}
\approx 20$ GHz at room temperature.

Room-temperature dielectric spectra of water, re-plotted from the
data of Refs. \cite{Franks, Hippel, Chaplin}, are shown in Fig. 1.
The $\varepsilon''(\omega)$ spectrum is dominated by an absorption
peak, accompanied by a step-like frequency dispersion in
$\varepsilon'(\omega)$. The step connects the high-frequency
dielectric constant, $\varepsilon_{\infty}$ = 5, with the
low-frequency (static) constant, $\varepsilon$(0) = 80. Let us note,
that $\varepsilon'(\omega)$ and $\varepsilon''(\omega)$ are
connected by the Kramers-Kronig relations, thus the high value of
the static permittivity, $\varepsilon$(0), is due to a large
integral intensity of the Debye absorption band in $\varepsilon''
(\omega)$.

The Debye relaxation in water, investigated experimentally and
theoretically in many details \cite{Franks, Hippel, Chaplin,
Buchner, Agmon1, Gaiduk, Sharma}, is surprisingly well described by
a simple relaxation formula:
\begin{equation} \label{Debye2}
\varepsilon'(\omega) = \varepsilon _{\infty} +\frac{\Delta
\varepsilon _{D}}{1+\omega ^{2}\tau _{D}^{2}} , \varepsilon
''(\omega) = \omega\tau\frac{\Delta\varepsilon _{D}}{1+\omega^{2}
\tau _{D}^{2}}.
\end{equation}
Here $\tau_{D}$ is the temperature-dependent relaxation time and
$\Delta \varepsilon_{D} = \varepsilon (0) - \varepsilon_{\infty}$ is
the contribution of the dielectric relaxation to the static
dielectric constant.

For a long time, the Debye's idea about the orientational motion of
water molecules has been exploited as the main microscopic mechanism
responsible for the static permittivity. Basically since its
introduction, it has been acknowledged, that the Debye model is
oversimplified \cite{Hippel}. Therefore, the model is being
permanently modified. The present-day models involve the dynamics of
protons and large molecular clusters \cite{Geissler, Bakker,
Buchner, Agmon1, Gaiduk, Sharma, Bukowski, Agmon2, Walbran,
Kornyshev}. Important is that in all these models, the geometry of
the water molecule is a substantial input parameter.

Here, we propose an interpretation of the dielectric spectrum of
water alternative to the Debye's approach. We argue that considering
exclusively the diffusive motion of protons in water is sufficient
for quantitative description of its dielectric spectrum at
frequencies lower than $10^{11}$ Hz.

\section{Remarks on existing experimental data}

Due to the reasons, which will become apparent in the course of the
article, our analysis is performed in terms of complex dynamical
conductivity, rather than the dielectric function. The complex
conductivity, $\sigma^{*} = Re(\sigma^{*}) + i Im(\sigma^{*})$, is
related to the dielectric constant \textit{via}:
$\varepsilon^{*}(\omega) = \varepsilon_{\infty} + i\sigma^{*}(\omega
)/(\varepsilon _{0}\omega )$, where $\varepsilon_{0}$ is the
free-space permittivity \cite{Landau}. Thus, the real part of
conductivity, $Re(\sigma^{*}) \equiv \sigma$, is merely the
imaginary part of permittivity multiplied by frequency:
$\sigma(\omega) = \omega\varepsilon_{0}\varepsilon ''(\omega)$.

\begin{figure}[]
\centering
\includegraphics[width=\columnwidth,clip]{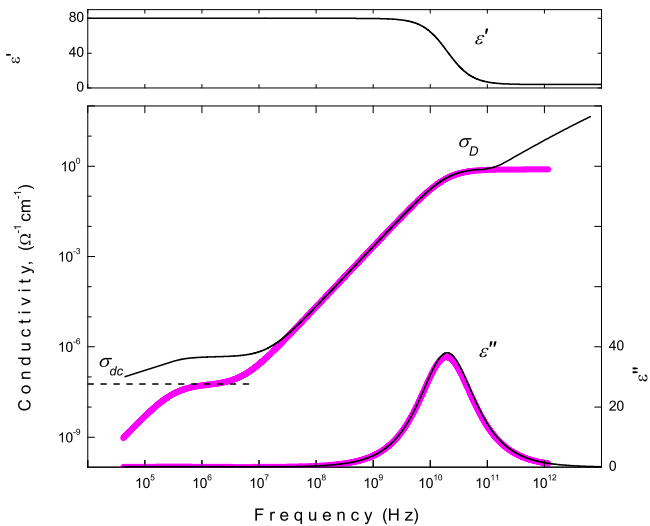}
\caption{(Color online) Panoramic dielectric response of water at
room temperature. Thin solid black lines are experimental data from
Ref. \cite{Hippel}. Dashed line is the frequency-independent (at $0
\leq \nu < 10^{7}$ Hz) conductivity according to Ref. \cite{Light,
Bockris}. Thick solid magenta lines are calculations with Eq.
\ref{sigma} and the parameters from Table 1. Top panel: real part of
the dielectric permittivity, $\varepsilon'(\omega)$. Bottom panel:
dynamical conductivity, $\sigma (\omega)$, left-hand scale, and
dielectric losses, $\varepsilon''(\omega)$, right-hand scale. }
\label{fig1}
\end{figure}

\begin{table}[b]
\caption{\label{table1} Experimental data from Refs. \cite{Hippel,
Light} used as input parameters in our model. All numbers are at
room temperature.}

\begin{ruledtabular}
\begin{tabular}{c|c|c|c}

 $\Delta\varepsilon_{D}$ & $\tau_{D}$ ($\equiv \tau_{2}$) & $\sigma_{dc} (\equiv \sigma_{1})$ & $\sigma_{2} \approx \sigma_{D}(\omega \rightarrow \infty)$ \\ \hline

 73 & $8.3 \times 10^{-12}$ s & $5.5  \times 10^{-6}$ $\Omega^{-1}$m$^{-1}$ & 78 $\Omega^{-1}$m$^{-1}$ \\

\end{tabular}
\end{ruledtabular}
\end{table}

The room-temperature $\sigma(\omega)$ spectrum of water is presented
in the main frame of Fig. 1 (left-hand scale). In order to be
specific, we use the data from Refs. \cite{Hippel, Light} for the
dielectric constant, $\Delta\varepsilon_{D}$, the relaxation time,
$\tau_{D}$, and the dc conductivity, $\sigma_{dc}$. The values of
these parameters for 25 $^\circ$C are listed in Table 1.

The most important, for further consideration, features of the
conductivity spectrum are the two well-distinguished plateaus,
$\sigma_{1}$ and $\sigma_{2}$, situated at $10^{6} - 10^{7}$ and
$10^{10} - 10^{11}$ Hz, respectively. The plateaus are connected by
a section, where $\sigma (\omega) \propto \omega^{s}$ with $s$ = 2.
At frequencies around the lower-frequency plateau, the available
experimental data differ from each other. On the one hand, there are
spectroscopic indications of a frequency-dependent conductivity in
this range \cite{Hippel} (thin black line in Fig. 1). On the other
hand, in electrochemistry the conductivity below $10^{7}$ Hz is
considered frequency-independent and equal to its dc value
\cite{Bockris} (dashed line in Fig. 1). Noteworthily, in the common
conductometry measurements, the working frequency is usually not
even mentioned \cite{Light}. Hereafter, we take $\sigma_{dc}$ as the
lower-frequency plateau value. Taking the data from Ref.
\cite{Hippel} instead, would only lead to minor quantitative
corrections.

\section{Arguments for our Model}

Being expressed in terms of $\sigma(\omega)$, the Debye relaxation
form (Eq. \ref{Debye2}) looks like:
\begin{equation} \label{Debye_sigma1}
\sigma_{D} \left(\omega \right)=\omega ^{2} \tau _{D} \varepsilon
_{0} \frac{\Delta \varepsilon _{D} }{1+\omega ^{2} \tau _{D}^{2} } .
\end{equation}
At high frequencies, $\sigma_{D}(\omega)$ is frequency-independent,
corresponding to the second plateau of the experimental spectrum:
\begin{equation} \label{sigma2}
\sigma_{2} \approx \sigma_{D}(\omega \rightarrow \infty) = \Delta
\varepsilon_{D} \varepsilon _{0}/\tau _{D}.
\end{equation}
Surely, there is no physical difference between the
representations in terms of $\varepsilon^{*}(\omega)$ and in terms
of $\sigma^{*}(\omega)$. However, we believe the conductivity
representation gives a tip for a fresh look on the dielectric
spectrum. Whereas the bell-shaped relaxation in
$\varepsilon''(\omega)$ is intuitively connected with the
orientational motion of the H$_{2}$O molecules, the
$\sigma(\omega)$ curve of Fig. \ref{fig1} hints to an alternative
mechanism, namely, to the proton diffusion.

In fact, the $\sigma $($\omega $) spectrum, given by Eq.
\ref{Debye_sigma1}, is consistent with acceleration-less motion of a
charge $q$ with mass $m$ in a parabolic potential, $\varphi(x) =
\kappa x^{2}/2$. The equation of motion for this charge is: $m\gamma
\mathop{x}\limits^{.} + \kappa x = qE$, where $\kappa$ is the spring
constant and $\gamma$ is the relaxation rate. Then, the mobility is
$\mu = q/(m\gamma)$, the diffusion coefficient is $D = k_{B}T\mu/q$,
and finally the conductivity is:
\begin{equation}\label{sigma_diffusion}
\sigma(\omega) = \varepsilon _{0} \frac{[q^{2}
N(k_{B}T)/D\kappa^{2} ]\omega ^{2} }{1+[(k_{B}T)^{2} /\kappa ^{2}
D^{2} ]\omega ^{2}},
\end{equation}
which coincides in spectrum shape with Eq. \ref{Debye_sigma1}.

Noteworthily, the diffusion of particles, interacting with
attractive centers, reveals such $D(\omega)$, that gives rise to the
same dispersion in conductivity as in our Eq. \ref{sigma_diffusion}
\cite{Stepisnik}.

The $\sigma(\omega)$ spectrum, consisting of two plateaus and a
$\sigma \propto \omega^{s}$ section in-between of them, is typical
for materials with high ionic conductivity, the superionics
\cite{Salomon}. The conductivity spectra of these materials have
been studied in great details. Their common feature is a suppression
of the dynamical conductivity at low frequencies due to localization
of diffusing particles in the minima of lattice potential and/or due
to interactions between the particles \cite{Salomon, Dieterich,
Dyre, Volkov}.

We believe, that in regard to its proton conductivity, water gives
all reasons to be compared with superionics. In accordance with the
modern concept, protons, H$^{+}$, and hydroxyl ions, OH$^{-}$, are
permanently generated (due to the self-ionization of H$_{2}$O
molecules \cite{Geissler, Bakker, Bukowski, Agmon2}) and recombined
in the volume of water.

Since free protons in water are not observed, they are considered to
localize after their birth on neighboring neutral H$_{2}$O molecules
(on femtosecond time scale). The excess proton converts the H$_{2}$O
molecule into a charged complex H$_{3}$O$^{+}$ with a positive
charge $q^{+}$, and leaves a ``hole", OH$^{-}$, with a negative
(twin) charge $q^{-}$. Subsequently, by a relay-race manner the
$q^{+}$ and $q^{-}$ charges wander diffusively over H$_{2}$O
molecules until they meet each other and recombine to produce a
neutral H$_{2}$O. The favorable-unfavorable molecular configurations
for the proton exchange are stochastically formed by the thermal
molecular motion \cite{Sharma, Walbran}.

The diffusion paths of the separated charges, from their birth to
recombination, are sketched in Fig. 2. The twin $q^{+}$ and $q^{-}$
charges are always in Coulomb field of each other. Therefore, in
their majority they do not go far from the places of their birth
(point 1 in Fig. 2). Instead, they recombine with their own twin
partner in a close vicinity of their birth places the [area with
characteristic size $\ell$, point 2 in Fig. 2]. Sometimes, however,
the ``twins" fail to meet each other and recombine with ``foreign"
partners on a much larger distance $L$ (points 3 and 4). Obviously,
the most probable foreign partner is a partner from the first
configuration sphere of the ionized molecules. Thus, effectively,
there are two recombination processes for charges in water, faster
and slower. We believe, that (similarly to superionics) the
characteristic lifetimes of the two processes reveal themselves as
characteristic knees in the conductivity spectrum.

\section{The Model}

The model outlined above can be described by a set of equations.
Let us designate the concentration, the average lifetime, and the
mean free path of the charges, involved in the slow and in the
fast processes, as $N_{1}$, $\tau_{1}$, $L$, and $N_{2}$,
$\tau_{2}$, $\ell$, respectively. $N_{1}$ and $N_{2}$, can also be
interpreted as the concentrations of the H$_{3}$O$^{+}$--OH$^{-}$
pairs participating in the slow and fast processes. Similarly,
$\tau_{1}$ and $\tau_{2}$ are the average life times of the
H$_{3}$O$^{+}$--OH$^{-}$ pairs, and $L$ and $\ell$ are their
characteristic sizes.

\begin{figure}[]
\centering
\includegraphics[width=\columnwidth,clip]{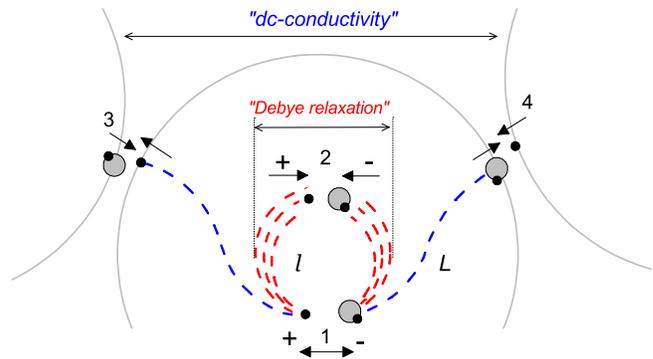}
\caption{(Color online) Schematic diagram of the two-scale proton
conductivity in water. The black dots are protons; the shadowed
small circles are oxygen atoms. The dashed lines show diffusion
paths of the charges $q^{+}$ and $q^{-}$, the single dashes
representing single proton hops. The large gray circles are
coordination spheres of ionized molecules. For further explanations,
see text.} \label{fig2}
\end{figure}

The basic assumption we made is that the lower and upper
conductivity plateaus in Fig. 1, $\sigma_{1}$ and $\sigma_{2}$,
correspond to diffusion of long- and short-living charges, the
latter (the fast) process giving rise to what is commonly refereed
as the Debye relaxation. Hence, the complete description of the
$\sigma (\omega)$ spectrum can be written as a sum of two terms:
\begin{equation}
\sigma \left(\omega \right)=\sigma _{1} \frac{\omega ^{2} \tau
_{1}^{2} }{1+\omega ^{2} \tau _{1}^{2} } +\sigma _{2} \frac{\omega
^{2} \tau _{2}^{2} }{1+\omega ^{2} \tau _{2}^{2} }. \label{sigma}
\end{equation}

The conductivity at each, the lower and the upper, plateau is
connected to the diffusion coefficient of charges, $D$, by the
Nernst-Einstein relation:
\begin{equation}
\sigma _{1} =2 C N_{1} D; \sigma _{2} =2 C N_{2} D,
\label{diffusion}
\end{equation}
where $C = q^{2}/k_{B}T$ and the coefficient 2 takes into account
the occurrence of positive and negative charges.

Using the Einstein-Smoluchowski formula, we can also connect $L$,
$\tau_{1}$, $\ell$, and $\tau_{2}$ to the diffusion coefficient:
\begin{equation}
D = \frac{\ell^{2}}{6\tau _{2}} = \frac{L^{2}}{6\tau _{1}}.
\label{es}
\end{equation}

Because both, slow and fast, processes span over all volume of
water, for the unit volume one can write:
\begin{equation}
1 = \frac{4\pi}{3} L^3 N_{1} = \frac{4\pi}{3} \ell^3 N_{2}.
\label{closepacked}
\end{equation}

Equations \ref{sigma} -- \ref{closepacked} constitute an equation
set, the analytical solution of which gives the following result:
\begin{equation}
N_{2} = 3 \left(4\pi\right)^{2}
\left(\frac{\sigma_{2}\tau_{2}}{C}\right)^{3}; \ell =
\frac{C}{4\pi\tau_{2}\sigma_{2}}; D = \frac{1}{6\tau_{2}^{3}}
\left(\frac{C}{4\pi \sigma_{2}}\right)^{2}; \label{result1}
\end{equation}
\begin{equation}
N_{1} =  N_{2} \times \frac{\sigma_{1}}{\sigma_{2}}; L = \ell
\times \left(\frac{\sigma_{2}}{\sigma_{1}}\right)^{1/3}; \tau_{1}
= \tau_{2} \times
\left(\frac{\sigma_{2}}{\sigma_{1}}\right)^{2/3}. \label{result2}
\end{equation}

The numerical values of these parameters, calculated using the
experimental data of Table 1, are presented in Table 2. With these
parameter values, Eq. \ref{sigma} comprehensively describes both,
the Debye relaxation and the dc conductivity (thick magenta line in
Fig. \ref{fig1}).

\begin{table}[t]
\caption{\label{table2} Room-temperature numerical values of the
parameters obtained from our model (Eqs. \ref{result1} --
\ref{result2}).}

\begin{ruledtabular}
\begin{tabular}{c|c|c}

 $N_{1}$ & $N_{2}$ & $D$ \\ \hline

 $3.8 \times 10^{19}$ m$^{-3}$ & $5.4 \times 10^{26}$ m$^{-3}$ & $1.2 \times 10^{-8}$ m$^{2}$/s \\ \hline \hline

 $L$ & $\ell$ & $\tau_{1}$\\ \hline

 0.18 $\mu$m & 0.76 nm & $5 \times 10^{-7}$ s

\end{tabular}
\end{ruledtabular}
\end{table}

The found parameters characterize a Brownian motion of thermally
activated charges $q^{+}$ and $q^{-}$ over the ``sea" of neutral
molecules, i.e. the drift currents. The mutual thermal motion of
neutral H$_{2}$O molecules is not relevant for our consideration.

\section{Consequences of the model and Conclusions}

1. \textit{Concentration of separated charges and life time of
H$_{2}$O molecules.} As one can see from Table 2, the concentration
of short-living H$_{3}$O$^{+}$--OH$^{-}$ pairs is huge, $N_{2}
\simeq 5 \times 10^{26}$ m$^{-3}$. Because $N_{2} \gg N_{1}$,
$N_{2}$ can be taken as the total concentration of the
H$_{3}$O$^{+}$--OH$^{-}$ pairs. Thus, roughly 1\% of all H$_{2}$O
molecules are ionized, the concentration of H$_{2}$O molecules being
$N_{0} = 3 \times 10^{28}$ m$^{-3}$. The found charge concentration
is by several orders of magnitude larger than the commonly accepted
value for neutral water, $10^{-7}$ mole/liter = $6 \times 10^{19}$
m$^{-3}$. Figuratively, water constantly boils with $\ell$-sized
H$_{3}$O$^{+}$--OH$^{-}$ pairs, which have a life time of $\tau_{2}
\simeq 10$ ps. This result agrees with what was reasoned in Ref.
\cite{Geissler} based on molecular-dynamics stimulations.

Our value for the size of H$_{3}$O$^{+}$--OH$^{-}$ pairs, $\ell =
0.76$ nm, correlates with the recent x-ray scattering results of
Ref. \cite{Nilsson}, where density fluctuations in water were
reportedly found on a comparable scale.

Because $N_{2} >> N_{1}$, the life time of a neural H$_{2}$O
molecule in thermodynamic equilibrium, $\tau_{0}$, can be
estimated from the following equation:
\begin{equation}
\frac{N_{2}}{\tau_{2}} = \frac{N_{0}}{\tau_{0}}. \label{tau0}
\end{equation}
From here, we obtain $\tau_{0} \simeq 1$ ns, which is 14 orders of
magnitude smaller than the ``standard" $10^{5}$ s.

2. \textit{Permittivity.} The Debye relaxation time changes its
meaning -- it is now the average life time of the short-living
separated charges (the H$_{3}$O$^{+}$--OH$^{-}$ pairs). For the
permittivity spectrum, by combining Eqs. \ref{sigma2}, \ref{sigma},
and \ref{diffusion}, we obtain:
\begin{equation}
\Delta \varepsilon_{D} = \frac{\sigma _{2}\tau_{2}}
{\varepsilon_{0}} = \frac{(q \times \ell)^{2}}{3kT} \frac
{N_{2}}{\varepsilon_{0}}, \label{polarizability}
\end{equation}
i.e., the pairs of the separated charges can be considered as
dipoles with the average dipole moment $p = q \times \ell$. The
polarizability (per unit volume) of these dipoles is $N_{2} \times
p^{2}/3kT$. Thus, Eq. \ref{polarizability} shows that the diffusion
of the charges to the average distance $\ell$ is the reason for the
step in the dielectric function, $\Delta \varepsilon_{D}$, at
frequencies around $1/2\pi\tau = 10^{10}$ Hz. The experimental value
of this step, $\Delta \varepsilon_{D} \approx 75$, is automatically
fulfilled in our model (through $N_{2}$ and $\ell$). This means that
the dominant contribution (more than 90\%) to the static dielectric
constant ($\varepsilon (0) \approx 80$) is provided by the
restricted-distance currents of separated charges, rather than by
orientational relaxation of intact H$_{2}$O dipoles.

Obviously, the orientational motion of the intact dipoles should
also reveal itself in the dielectric spectrum. We believe, a good
candidate for this is the Debye-like bands, found in the
measurements at frequencies higher than $10^{11}$ Hz \cite{Ronne,
Moller}. These higher-frequency relaxation processes are also seen
as the upturn of the experimental conductivity at the highest
frequencies in Fig. 1.

The slow recombination process with the characteristic time
$\tau_{1}$ also provides a contribution to the static dielectric
constant. According to Eq. \ref{sigma2}, this contribution is
however very small, $\Delta\varepsilon_{1} < 1$.

3. \textit{Diffusion coefficient.} In our consideration, the
diffusion coefficient, $D = 1.2 \times 10^{-8}$ m$^{2}$/s is related
to the relay-race diffusion of charges ($q^{+}$ and $q^{-}$), not to
diffusion of a tagged proton. In literature, however, this value
(more accurately, $9.3 \times 10^{-9}$ m$^{2}$/s) is commonly
accepted as the diffusion coefficient of protons \cite{Bockris} and
considered to be ``anomalously" high. According to our findings, the
real Brownian diffusion coefficient of a tagged proton, $D^{H+}$, is
100 times smaller: $D^{H+} = (a^{2}/6\tau_{0}) \approx 10^{-10}$
m$^{2}$/s, where $a$ = 2.5~{\AA} is the distance between the
H$_{2}$O molecule centers \cite{Agmon2}.

It is worth noting here, that if one substitutes the proton
diffusion coefficient, $D^{H+}$, and the total proton concentration,
$2N_{0}$, into the Nernst-Einstein equation (Eq. \ref{diffusion}),
one gets the upper limit for the proton conductivity in water:
$\sigma = 4 C N_{0} D^{H+} = 74$ $\Omega^{-1}$m$^{-1}$, that is
precisely the value of the higher-frequency plateau in the
$\sigma(\omega)$ spectrum, $\sigma_{2}$.

4. \textit{dc conductivity and relevance to pH index.} The found
value for $N_{1}$ ($3.8 \times 10^{19}$ m$^{-3}$) is practically
equal to the commonly accepted concentration of ``free" protons ($6
\times 10^{19}$ m$^{-3}$), which provides pH = 7 in neutral water.
This result shows that the common dc conductometric methods detect
only those ``survived" protons (of concentration $N_{1}$), which are
involved in the slow recombination process, while the short-living
protons (concentration $N_{2}$) are not detectable in these
measurements. The occurrence of the short-living protons in water
requires introducing a new, ``fast", dissociation constant,
$K_{W2}$, in addition to the common (``slow") $K_{W1}$, related to
the pH index. Since both, long- and short-living, protons are
chemically active, the conventional conception of the pH index could
probably be revised in such a way that the fast $K_{W2}$ is also
taken into account.

Summarizing, we have found that the dielectric spectrum of liquid
water at frequencies below $10^{11}$ Hz can be entirely understood
in terms of proton diffusion, resulting from self-dissociation of
H$_{2}$O molecules. No long-living geometric structures, created by
the water molecules, are needed to be included in this
consideration. We believe that many other physical properties of
water are also determined mostly by the ability of H$_{2}$O
molecules to dissociate rather than to form any sorts of geometric
structures.

We are grateful to S. D. Zakharov and G. M. Zhidomirov for useful
discussions.

\end{document}